\documentclass[aps]{revtex4}

\begin{document}
\draft \preprint{v21-ARL Tech Report}

\title{Stress in Rotating Disks and Cylinders}
\author{Thomas B. Bahder}
\address{U. S. Army Research Laboratory \\
2800 Powder Mill Road \\
Adelphi, Maryland, USA  20783-1197}

\date{\today}

\begin{abstract}
The solution of the classic problem of stress in a  rotating
elastic disk or cylinder, as solved in standard texts on
elasticity theory, has two features: dynamical equations are used
that are valid only in an inertial frame of reference, and
quadratic terms are dropped in displacement gradient in the
definition of the strain. I show that, in an inertial frame of
reference where the dynamical equations are valid, it is incorrect
to drop the quadratic terms because they are as large as the
linear terms that are kept. I provide an alternate formulation of
the problem by transforming the dynamical equations to a
corotating frame of reference of the disk/cylinder, where dropping
the quadratic terms in displacement gradient is justified. The
analysis shows that the classic textbook derivation of stress and
strain must be interpreted as being carried out in the corotating
frame of the medium.

\end{abstract}

\maketitle

\section{Introduction}

The problem of stresses in rotating disks and cylinders is
important in practical applications to rotating machinery, such as
turbines and generators,  and wherever large rotational speeds are
used. The textbook problem of stresses in elastic rotating disks
and cylinders, using the assumption of plane strain or plane
stress, is published in classic texts, such as
Love~\cite{Love1944}, Landau and Lifshitz~\cite{LLelasticity1970},
Nadai~\cite{Nadai1950}, Sechler~\cite{Sechler1952}, Timoshenko and
Goodier~\cite{Timoshenko1970}, and Volterra and
Gaines~\cite{VolterraGaines1971}. The standard approach presented
in these texts has two characteristic features:
\begin{enumerate}
\item Newton's second law of motion is applied in an inertial frame of
reference to derive dynamical equations for the continuum (see Eq.~(\ref{stressEq}) below), and
\item quadratic terms in displacement gradient are dropped in the definition of the strain
tensor (see Eq.(\ref{strainDisplacement}) below).
\end{enumerate}
In this paper, I show that, for a rotating elastic body, the
second feature of the solution is inconsistent with the first:
dropping the quadratic terms in the displacement gradient is an
unjustified approximation in an inertial frame of reference. In
what follows, I refer to the method that is employed in
Ref.~\cite{Love1944,LLelasticity1970,Nadai1950,Sechler1952,Timoshenko1970,VolterraGaines1971}
as `the standard method', and for brevity, I will refer to a
cylinder as a generalization of both a disk and a cylinder.

The classic problem of stress in an elastic rotating cylinder is
complex because the undeformed reference state of the body is the
non-rotating state.  The deformed state is one of steady-state
rotation.  The analysis of the problem must connect the
non-rotating reference state to the rotating stressed/strained
state. These two states are typically connected by large angles of
rotation.  When large angles of rotation are present, the
quadratic terms in the displacement gradient cannot be dropped (in
an inertial frame of reference) in the definition of the
strain~\cite{Dienes1979,Dienes1986,NaghdiEtAl}. The problem of
stress analysis when large-angle rotations are present is well
known and has been discussed by a number of authors in general
contexts, see for example~\cite{Dienes1979,Dienes1986,NaghdiEtAl}.
However, large-angle rotations in the problem of a rotating
elastic cylinder have not been dealt with in a technically correct
manner, because quadratic strain gradient terms are incorrectly
dropped in the `standard
method'~\cite{Love1944,LLelasticity1970,Nadai1950,Sechler1952,Timoshenko1970,VolterraGaines1971}.

In this work, I formulate the elastic problem of a rotating
cylinder in a  frame of reference that is corotating with the
material.  In this corotating frame, the quadratic terms in the
displacement gradient can be dropped, and the resulting
differential equations are linear and can be solved.

In section II, I review the `standard method' of solution used in
Ref.~\cite{Love1944,LLelasticity1970,Nadai1950,Sechler1952,Timoshenko1970,VolterraGaines1971}
and show that for a rotating cylinder the displacement gradient in
an inertial frame of reference is of order unity, and therefore
quadratic terms (in strain tensor definition) cannot be dropped
when compared to the linear terms. Section III contains the bulk
of the analysis. I describe the corotating systems of coordinates
and the transformation of the velocity field to the corotating
frame. I use the velocity transformation rules to transform the
dynamical Eq.~(\ref{stressEq}) from the inertial frame to the
corotating frame (see Eq.~(\ref{corotatingStresEq}) or
(\ref{CylindricalCorotatingStresEq})), where extra terms arise
known as the centrifugal acceleration and the coriolis
acceleration.  In section IV, I write the explicit component
equations for stress (in cylindrical coordinates) for the rotating
elastic cylinder in its corotating frame.  To display the
resulting solution concretely, I derive the well-known formula for
the stress in the rotating cylinder for the case of plane stress,
as computed in the corotating frame. Stress is an objective
tensor, i.e., stress is independent of observer
motion~\cite{Eringen1962,Narasimhan1992}, so the physical meaning
of stress in the corotating frame is the same as in the inertial
frame. Therefore, the stress field components in the corotating
frame are equal to the stress field components in the inertial
frame, see Eq.~(\ref{stressTransformation}).

\section{Standard Solution Method}

In the `standard
method'~\cite{Love1944,LLelasticity1970,Nadai1950,Sechler1952,Timoshenko1970,VolterraGaines1971},
the stress analysis of elastic rotating cylinders starts with the
dynamical equations, which, in generalized curvilinear coordinates
are given by~\cite{LLelasticity1970,Eringen1962,Narasimhan1992}
\begin{equation}
\sigma^{k j}_{~~ ;j} + \rho f^k = \rho \, a^k \label{stressEq}
\end{equation}
where $\sigma^{k j}$ are the contravariant components of the
stress tensor, $f^k$ is the vector body force, and $a^k$ is the
acceleration vector. In Eq.~(\ref{stressEq}), repeated indices are
summed and the semicolon indicates covariant differentiation with
respect to the coordinates. Expressed in terms of the velocity
field in spatial coordinates, the acceleration is given
by~\cite{Eringen1962,Narasimhan1992}
\begin{equation}
a^k = \frac{\partial v^k}{\partial t} + v^j v^k_{~ ;j}
\label{accelDef}
\end{equation}
where  $v^a$ is the velocity field, and the semicolon indicates
covariant differentiation with respect to the coordinates, and
$v^j v^k_{~ ;j}$ is called the convective term.  In
Eq.~(\ref{stressEq}), the stress  $\sigma^{kj}$,  acceleration
$a^k$, and body force $f^k$, are generally time dependent.
Equation~(\ref{stressEq}) is derived by applying Newton's second
law of motion to an element of the medium.  Newton's second law is
valid only in an inertial frame of reference, and consequently the
validity of Eq.~(\ref{stressEq}) is limited to inertial frames of
reference.

In the `standard  method' of solution, Eq.~(\ref{stressEq}) is
applied by invoking an ``effective body force", of magnitude equal
to the centrifugal force in the rotating frame. In the inertial
frame, there is actually no effective force (such as Coriolis or
cetrifugal force). For the case of a body rotating about its
principle axis, a more careful determination of the terms $ f^k -
a^k$ in Eq.~(\ref{stressEq}) comes from setting the body force to
zero (or setting equal to some applied force) and computing the
material acceleration $a^k$ for a given body motion. For a rigid
body, or a uniform density elastic cylinder that is rotating about
its axis of symmetry at a constant angular velocity $\omega_o$,
the Cartesian velocity field components are: $v^1=-\omega_o \, y$,
$v^2= \omega_o \, x$, and $v^3= 0$, where superscripts 1,2,3
indicate components on the Cartesian basis vectors associated with
the $x$,$y$,$z$-axes (in the inertial frame).  Corresponding to
this velocity field,  the cylindrical components of the
acceleration field are given by
\begin{equation}
\bar{a}^k =  \frac{\partial \bar{v}^k}{\partial t} + \bar{v}^b \bar{v}^k_{~ ;b}
 = \left(-r \omega_o^2 , 0 , 0 \right)
\label{accel}
\end{equation}
where I have chosen the z-axis as the symmetry axis and the bar
over the components indicates that they are in the inertial frame
of reference in cylindrical coordinates.  For the case where there
are no body forces,  with the acceleration in Eq.\ (\ref{accel}),
 Eq.\ (\ref{stressEq}) in cylindrical coordinates leads to the
three equations
\begin{eqnarray}
\bar{\sigma}^{11}_{~~ ,1} + \bar{\sigma}^{12}_{~~ ,2} + \bar{\sigma}^{13}_{~~ ,3} +
\frac{\bar{\sigma}^{11}}{r}- r \bar{\sigma}^{22} & = &  - \rho r \omega_o^2   \label{stress1} \\
\bar{\sigma}^{12}_{~~ ,1} + \bar{\sigma}^{22}_{~~ ,2} + \bar{\sigma}^{23}_{~~ ,3} +
\frac{3}{r} \bar{\sigma}^{12}  & = &   0   \label{stress2} \\
\bar{\sigma}^{13}_{~~ ,1} + \bar{\sigma}^{23}_{~~ ,2} + \bar{\sigma}^{33}_{~~ ,3} +
\frac{\bar{\sigma}^{13}}{r}   & = &  0  \label{stress3}
\end{eqnarray}
where the superscripts 1,2,3 enumerate tensor components on the
$r,\phi,z$ coordinate basis vectors respectively,  in cylindrical
coordinates and the commas indicate partial differentiation with
respect to these coordinates.

For steady rotation at a uniform angular velocity $\omega_o$, and
assuming the absence of elastic waves, there is rotational
symmetry about the $z$-axis so the stress components do not depend
on azimuthal angle $\phi$. Therefore,  all derivatives with
respect to $\phi$ are zero, leading to the equations:
\begin{eqnarray}
\bar{\sigma}^{11}_{~~ ,1} + \bar{\sigma}^{13}_{~~ ,3} +
\frac{\bar{\sigma}^{11}}{r}- r \bar{\sigma}^{22} & = &  - \rho r \omega_o^2   \label{stress4} \\
\bar{\sigma}^{12}_{~~ ,1}  + \bar{\sigma}^{23}_{~~ ,3} + \frac{3}{r} \bar{\sigma}^{12}  & = &   0   \label{stress5} \\
\bar{\sigma}^{13}_{~~ ,1} + \bar{\sigma}^{33}_{~~ ,3} + \frac{\bar{\sigma}^{13}}{r}   & = &  0  \label{stress6}
\end{eqnarray}

I introduce physical components of stress, $\sigma^{rr}$,
$\sigma^{\phi \phi}$,$\sigma^{z z}$, $\sigma^{r \phi}$, $\sigma^{r
z}$, and $\sigma^{\phi z}$, with units of force per unit area and
which are related to the tensor components $\bar{\sigma}^{11}$,
$\bar{\sigma}^{22}$, $\bar{\sigma}^{33}$, $\bar{\sigma}^{12}$,
$\bar{\sigma}^{13}$,  and $\bar{\sigma}^{23}$,
by~\cite{Eringen1962,Narasimhan1992}
\begin{eqnarray}
\bar{\sigma}^{r r} & = & \bar{\sigma}^{11}, \label{pc1} \\
\bar{\sigma}^{\phi \phi} & = & r^2 \bar{\sigma}^{22}  \label{pc2}\\
\bar{\sigma}^{z z} & = & \bar{\sigma}^{33}  \label{pc3} \\
\bar{\sigma}^{r \phi} & = & r \bar{\sigma}^{12} \label{pc4} \\
\bar{\sigma}^{r z} & = & \bar{\sigma}^{13} \label{pc5} \\
\bar{\sigma}^{\phi z} & = & r \bar{\sigma}^{23} \label{pc6}
\end{eqnarray}
Expressing  Eq.~(\ref{stress4})--(\ref{stress6}) in terms of the
physical components, I obtain the well-known equations valid in an
inertial frame of
reference~\cite{Love1944,LLelasticity1970,Nadai1950,Sechler1952,Timoshenko1970,VolterraGaines1971},
\begin{eqnarray}
\frac{\partial \bar{\sigma}^{r r}}{\partial r} + \frac{\partial \bar{\sigma}^{r z}}{\partial z} +
\frac{\bar{\sigma}^{rr} - \bar{\sigma}^{\phi \phi}}{r} & = &  - \rho r \omega_o^2   \label{st4} \\
\frac{\partial}{\partial r} \left( \frac{1}{r} \bar{\sigma}^{r \phi} \right)
+ \frac{1}{r} \frac{\partial \bar{\sigma}^{\phi z}}{\partial z}  + \frac{3}{r^2} \bar{\sigma}^{r \phi}  & = &   0   \label{st5} \\
\frac{\partial \bar{\sigma}^{r z}}{\partial r}  + \frac{\partial \bar{\sigma}^{z z}}{\partial z}  + \frac{\bar{\sigma}^{r z}}{r}   & = &  0  \label{st6}
\end{eqnarray}
Note that Eqs.~(\ref{st4})--(\ref{st6}) have been derived using
Newtons's second law, and so they are valid only in an inertial
frame of reference.  In particular, Eqs.~(\ref{st4})--(\ref{st6})
are not valid in a rotating frame of reference.

When a rotating disk or cylinder is analyzed, the assumption of
plane stress or plane strain is often made.  In both cases,
stresses must be related to strains by constitutive equations. For
the simplest case of a homogeneous, isotropic, perfectly elastic
body, the constitutive equations in curvilinear coordinates in an
inertial frame can be written
as~\cite{LLelasticity1970,Eringen1962,Narasimhan1992}
\begin{equation}
\sigma^{ik} = \lambda \, e \, g^{ik} + 2 \, \mu \, e^{ik}
 \label{constitutiveEq}
\end{equation}
where $\lambda$ and $\mu$ are the Lam\'{e} material constants,
$e^{ik}$ are the contravariant strain tensor components, $e=e^{\,
~a}_a$ is the contraction of the strain tensor, and $g^{ik}$ are
the contravariant metric tensor components.

The Eulerian strain tensor $e_{ik}$ is related to the displacement
field $u^i$ by~\cite{Eringen1962,Narasimhan1992}
\begin{equation}
e_{ik} = \frac{1}{2} \left( u_{i;k} + u_{k;i} + u_{m;i} u^m_{~~ ;
k} \right) \label{strainDisplacement}
\end{equation}
In the `standard  method' of
solution~\cite{Love1944,LLelasticity1970,Nadai1950,Sechler1952,Timoshenko1970,VolterraGaines1971},
the quadratic terms $u_{m;i} u^m_{~~ ; k}$ are dropped, which
leads to linear equations that can be solved (for example, by
using the Airy stress function~\cite{LinearizationComment}).

However, dropping the quadratic terms in Eq.\
(\ref{strainDisplacement}) is not justified for a rotating body
because these (dimensionless) terms $u_{m;i}$ are of order unity.
To prove this assertion, it is sufficient to consider the limiting
case of a rigid body in steady-state rotation at constant angular
speed $\omega_o$. The deformation mapping function gives the
coordinates $z^k$ (here taken to be Cartesian) of a particle at
time $t$ in terms of the particle's coordinates $Z^k$ in some
reference state (configuration) at time $t=t_o$:
\begin{equation}\label{defMapFcn}
z^k = z^k(Z^m,t)
\end{equation}
so that $z^k(Z^m,t_o)=Z^k$.  The
deformation mapping function has an inverse, which I quote here
for later reference
\begin{equation}\label{defMapFcnInverse}
Z^m = Z^m(z^k,t)
\end{equation}
Both $z^k$ and  $Z^m$ refer to the same Cartesian coordinate
system. The coordinates of a  particle initially at $Z^k$ at $t=t_o=0$
rotating about the $z$-axis are given by the deformation mapping
function
\begin{equation}\label{DefMappingFunction}
z^k = R^k_{~ \,m} Z^m
\end{equation}
where the orthogonal matrix $R^k_{\, ~ m}$ is given by
\begin{equation}\label{Rmatrix}
 R^k_{\, ~m} = \left( \begin{array}{ccc}
   \cos \omega_o t & \sin \omega_o t & 0 \\
    -\sin \omega_o t & \cos \omega_o t  & 0 \\
   0 & 0 & 1
\end{array} \right)
\end{equation}
The displacement vector field for this deformation mapping
function is given by~\cite{Spencer1980}
\begin{equation}\label{displacementField}
{\bf u} = u^m {\bf I}_m = (z^m - Z^m ) {\bf I}_m = (\delta^m_k -
\bar{R}^m_{~ \, k} )z^k {\bf I}_m
\end{equation}
where ${\bf I}_m$ are the unit Cartesian basis vectors and
$\bar{R}^m_{\, ~k}$ is the transpose matrix that satisfies
\begin{equation}\label{InverseRelation}
  \bar{R}^m_{~~k} R^k_{~~l} = \delta^m_{l}
\end{equation}
where $\delta^m_l=+1$ if $m=l$ and $0$ if $m \ne l$. Therefore,
from Eq.~(\ref{displacementField}), it is clear that gradients of
displacement $u_{m;k}$ appearing in Eq~(\ref{strainDisplacement})
are of order unity and therefore the quadratic terms $u_{m;i}
u^m_{~~ ; k}$ cannot be dropped because they are not small.  More
specifically, the dropped terms in Eq.~(\ref{strainDisplacement})
vary in time between -2 and 0 (in Cartesian components):
\begin{equation}\label{DroppedTerms}
\frac{1}{2} \, u_{m;i} \, u^{m}_{~~;k} = \left(
\begin{array}{ccc}
   -1 + \cos \omega_o t &   0                &     0 \\
    0                 & -1 + \cos \omega_o t  &   0 \\
   0 & 0 & 0
\end{array} \right)
\end{equation}
The above calculation was done for a rigid body, but clearly, a
similar error is introduced for elastic bodies. Therefore, in
general, for a rotating elastic body, the quadratic terms in
displacement gradients in Eq.~(\ref{strainDisplacement}) cannot be
dropped~\cite{FOOTNOTE-1}.

In cylindrical components, the relation between the physical
components  (see Ref.~\cite{Eringen1962,Narasimhan1992}) of
strain, $\bar{e}_{rr}$ and $\bar{e}_{\phi \phi}$, and physical
components of the displacement field, $(u_r,u_\phi,u_z)$, is given
by
\begin{eqnarray}
\bar{e}_{rr} & = &  \frac{\partial u_r}{\partial r} - \frac{1}{2}\, \left[
\left( \frac{\partial u_r}{\partial r}\right)^2 +
\left( \frac{\partial u_\phi}{\partial r} \right)^2 +
\left( \frac{\partial u_z}{\partial r}\right)^2  \right]  \label{strainU1} \\
\bar{e}_{\phi \phi} & = &  \frac{u_r}{ r} - \frac{1}{2 r^2}\,
\left[  (u_r)^2 + (u_\phi)^2 \right]  \label{strainU2}
\end{eqnarray}
(I use a bar over  $\bar{e}_{rr}$ and $\bar{e}_{\phi \phi}$ to
indicate that these are cylindrical components, and indices $rr$
and $\phi\phi$ (as distict from 11 and 22) to indicate that these
are physical components and not tensor components. See the
Appendix and Table I and II for notation conventions.)  In
Eq.~(\ref{strainU1}) and (\ref{strainU2}), I have assumed that
there is no dependence on $\phi$ and $z$, so I have set
derivatives with respect to these variables to zero.

In the `standard method' of solving for the stress in a rotating
cylinder~\cite{Love1944,LLelasticity1970,Nadai1950,Sechler1952,Timoshenko1970,VolterraGaines1971},
the quadratic terms in Eq.~(\ref{strainU1}) and (\ref{strainU2})
are incorrectly dropped.

The straight forward approach to correctly studying the stresses
in a rotating disk or cylinder, involves keeping the quadratic
terms in displacement gradient in Eq.~(\ref{strainDisplacement}).
However, this approach does not appear promising because it leads
to insoluble nonlinear differential equations. In the next
section, I approach the problem by using a transformation to a
corotating frame of reference, in which dropping the quadratic
terms can be justified for moderate angular velocity of rotation
$\omega_o$.

\section{Transformation to the Rotating Frame}

As discussed in the introduction, the problem of an elastic
rotating cylinder is complicated because the unstressed/unstrained
reference state is the non-rotating state, while the stressed
(strained) state is rotating, and these two states are typically
related by a large (time-dependent) angle. The analysis of the
rotating disk or cylinder must relate the stresses in the rotating
state to the reference configuration, which I take to be the
non-rotating state.  I define a transformation from an inertial
frame of reference to the corotating frame of reference of the
cylinder. This transformation provides a relation between the
rotating stressed state and the non-rotating reference
configuration.

\subsection{Coordinate Systems}

Starting from an inertial frame of reference, $S$, defined by the
Cartesian coordinates $z^k=(x,y,z)$, I make a transformation to a
rotating frame of reference $S^\prime$. The rotating frame will be
corotating with the cylinder so that in this frame $S^\prime$ the
azimuthal velocity field will be zero at all times.  The
transformation from an inertial system of coordinates to a
rotating system of coordinates is most simply done using Cartesian
coordinates.  On the other hand, the assumed cylindrical symmetry
of the problem begs for use of cylindrical coordinates. Hence I
will make use of four systems of coordinates. In the inertial
frame $S$, I have two systems of coordinates: a Cartesian system
of coordinates $z^k=z_k=(x,y,z)$, and a cylindrical system of
coordinates $x^i=(r,\phi,z)$.  In the corotating frame of
reference, $S^\prime$, I have a Cartesian system of coordinates
$z^{\prime k}=(x^\prime,y^\prime,z^\prime)$ and a cylindrical
coordinate system $x^{\prime i}=(r^\prime,\phi^\prime,z^\prime)$.
These coordinates are summarized in Table~\ref{coordinates} and
the Appendix.  I also introduce notation for tensor components in
each of the four coordinate systems, see
Table~\ref{TensorComponents} . The Cartesian components of the
stress tensor in the inertial frame $S$ will be denoted by
$\sigma^{ik}$.  In the same inertial frame $S$, the cylindrical
components of stress will be $\bar{\sigma}^{ik}$. The Cartesian
components of stress in the corotating frame $S^\prime$ will have
a prime, $\sigma^{\prime ik}$.  In this same corotating frame,
$S^\prime$, the cylindrical components of stress will be denoted
by using a tilde, $\tilde{\sigma}^{ik}$.

From the vantage point of an inertial frame of reference, $S$,
with Cartesian coordinates $z^k$, consider a cylinder whose
symmetry axis is aligned and colocated with the coordinate z-axis.
At time $t=-\infty$, take the cylinder to be non-rotating.  Now
assume that in the distant past, around the time $t\sim -T$, the
cylinder begins a slow angular acceleration lasting a long time,
on the order of $1/\epsilon$, where $1/\epsilon << T$. An example
of such an angular acceleration function is
\begin{equation}\label{angularAcceleration}
\omega(t) = \frac{1}{2} \omega_o \left[ 1+\tanh( \epsilon(t+T))
\right]
\end{equation}
where I assume that $\tau << 1/\epsilon << T$ and $\tau$ is the
longest time constant in the problem.  This inequality states that
the acceleration occurs slowly, $\tau << 1/\epsilon$, slower than
any time scale in the problem, and that this acceleration occurs
in the distant past, $1/\epsilon << T$, so that at $t=0$, I have a
steady-state situation of a cylinder rotating at constant angular
speed $\omega_o$. By slowly accelerating the cylinder, I avoid
introducing modes of vibration.  As the cylinder's angular
velocity increases from $t=-\infty$, each particle comprising the
cylinder moves along a spiral trajectory (with increasing radius).
From the point of view of the inertial frame $S$, the stresses on
a given element of the medium (particle) are such that they cause
the particle to experience an acceleration, moving along the
spiral path.  At $t=0$, the cylinder has achieved its maximum
angular velocity $\omega_o$. Due to the assumption of a perfectly
elastic medium, at $t=0$ the velocity field has zero radial
component; all particles of the cylinder are moving azimuthally
(in a plane perpendicular to the $z$-axis with zero radial
component).   The velocity field is that of a rigid body and the
acceleration field is given by Eq.~(\ref{accel}).

Now I introduce the corotating frame of reference, $S^\prime$,
with the Cartesian coordinates $z^{\prime}_k$, whose angular
velocity of rotation is equal to that of the cylinder at all
times. The coordinates $z_k \, (\equiv \, z^k)$ and
$z^{\prime}_{k} \, (\equiv \, z^{\prime \, k})$ are related by
\begin{equation}\label{rotatingFrameTransformation}
z^\prime_i = A_{ik}(t) \, \, z_k
\end{equation}
where the time dependent matrix $A_{ik}(t)$ is given by
\begin{equation}\label{matrixA}
A_{ik}(t)= \left(
\begin{array}{ccc}
  \cos (\theta-\theta_o)  & \sin (\theta-\theta_o) &  0 \\
  -\sin (\theta-\theta_o) & \cos (\theta-\theta_o) & 0 \\
  0 & 0 & 1
\end{array}
\right)
\end{equation}
where $\theta$ is a function of time given by the integral of
$\omega(t)$:
\begin{equation}\label{thetaDef}
\theta(t) = \frac{\omega_o}{2 \epsilon}
\left[ \epsilon (t+T) + \log\cosh(\epsilon t +\epsilon T) + \log 2 \right]
\end{equation}
and $\theta(0)=\theta_o$.

By construction, in the corotating frame $S^\prime$ the particles
comprising the material are not rotating about the
$z^\prime$-axis; there is zero azimuthal component of the velocity
field at all times.   As the angular velocity $\omega(t)$
increases from $t=-\infty$, each particle comprising the cylinder
experiences an increasing effective centrifugal force that
displaces the particle to a larger radius. In this rotating frame
$S^\prime$, there will (in general) also be a Coriolis force.
However, in $S^\prime$, for moderate angular speed $\omega_o$, the
strain $e_{ik}$ will be small, and the gradients of the
displacement field will also be small. Consequently, dropping the
quadratic terms $u_{m;i} u^m_{~~ ; k}$ in
Eq.~(\ref{strainDisplacement}) will provide a good approximation
to $e_{ik}$.

Note that the transformation that relates cylindrical components
in inertial frame $S$ and rotating frame $S^\prime$ is given by
the identity matrix
\begin{equation}\label{SandSprime}
\frac{\partial x^i}{\partial x^{\prime \, k}} = \,
\frac{\partial x^i}{\partial z^a} \,
\frac{\partial z^a}{\partial z^{\prime \, m}} \,
\frac{\partial z^{\prime \,m}}{\partial x^{\prime \, k}} \, = \,
\delta^i_k
\end{equation}
Furthermore, the time dependent transformation between inertial
cylindrical coordinates $x^i=(r,\phi,z)$ in $S$ and corotating
cylindrical coordinates
 $x^{\prime \, i}=(r^\prime,\phi^\prime,z^\prime)$ in $S^\prime$,
 is given by
\begin{eqnarray}
r^\prime & = & r \nonumber \\
\phi^\prime & = & \phi - \left( \theta(t)-\theta_o  \right) \label{CylTrans} \\
 z^\prime & = & z   \nonumber
\end{eqnarray}
where $\theta(t)$ is given by Eq.~(\ref{thetaDef}).
The relation between cylindrical stress components
$\bar{\sigma}^{ik}$ in the inertial frame $S$ and
cylindrical stress components $\tilde{\sigma}^{ik}$
in the corotating frame $S^\prime$ is
\begin{equation}\label{stressTransformation}
\bar{\sigma}^{ik} (x^n) = \frac{\partial x^i}{\partial x^{\prime \, a}}
\frac{\partial x^k}{\partial x^{\prime \, b}} \,
\tilde{\sigma}^{ab} (x^{\prime \, n}) \,
= \, \tilde{\sigma}^{ik} (x^{\prime \, n})
\end{equation}
Of course the transformation in Eq.~(\ref{stressTransformation})
must be used so that the components are referring to the same
physical point in space having coordinates $x^n=(r,\phi,z)$ and
$x^{\prime \, n} = (r^\prime,\phi^\prime,z^\prime)$, where the
time-dependent relation between $x^n$ and $x^{\prime \, n}$ is
given by Eq.~(\ref{CylTrans}).

The Eulerian strain tensor $e_{ij}$ transforms in a more
complicated manner. A general deformation is given by
\begin{equation}\label{DeformGeneral}
x^i=x^i(X^k,t)
\end{equation}
where a particle at time $t=t_o$ in the reference configuration
has (curvilinear) coordinates $X^k$, and in the deformed state at
time $t$ the particle has coordinates $x^i$ (in the same
curvilinear coordinate system).  The Eulerian strain tensor
$e_{ij}(X,x)$ depends on two points: $X$ in the reference
configuration and $x$ in the deformed state.  Consequently, under
a general coordinate transformation to a moving frame, $x^i
\rightarrow x^{\prime \, i}=h^i(x^k,t)$, the Eulerian strain is a
two-point tensor, which transforms as a second rank tensor under
transformation of deformed coordinates
\begin{equation}\label{x-Transformation}
x^i \rightarrow x^{\prime \, i} = h^i(x^{k},t)
\end{equation}
and transforms as a
scalar under transformation of reference state coordinates
\begin{equation}\label{X-Transformation}
X^i \rightarrow X^{\prime \, i} = h^i(X^k,t_o)
\end{equation}
so
that~\cite{Murnaghan1937,GambiEtAl1989,Bahder-strainTransformation2000}
\begin{equation}\label{strainTransformationGeneral}
\bar{e}_{mn}(X,x)= \tilde{e}_{ik}(X^\prime,x^\prime) \,
\frac{\partial x^i}{\partial x^{\prime \, m}} \, \frac{\partial
x^k}{\partial x^{\prime \, n}} = \tilde{e}_{mn}(X^\prime,x^\prime)
\end{equation}
where $x$ and $x^\prime$, and $X$ and $X^\prime$, are related by
Eq.~(\ref{x-Transformation}) and (\ref{X-Transformation}), and I
used Eq.~(\ref{SandSprime}). Therefore, the cylindrical components
of strain in the inertial frame $S$, $\bar{e}_{mn}(X,x)$, are
equal to the cylindrical components of strain in the rotating
frame $S^\prime$, $\tilde{e}_{ik}(X^\prime,x^\prime)$. Finally,
since I assume rotational symmetry about the $z$-axis (and
$z^\prime$-axis) so that all physical quantities have no
dependence on $\phi$ or $\Phi$, which are the azimuthal
coordinates of the point in the deformed state,
$x=x^i=(r,\phi,z)$, and coordinates of the point in the reference
configuration, $X=X^i=(R,\Phi,Z)$. Because of the nature of the
transformation to the rotating frame in Eq.~(\ref{CylTrans}),
Eq.~(\ref{strainTransformationGeneral}) can be used with
$r^\prime=r$, $R^\prime=R$,  $z^\prime = z$, and $Z^\prime = Z$.
Therefore, the tensor components in the corotating frame are
identical to the components in the inertial frame.

\subsection{Transformation of Lagrangean and Eulerian Velocities}

The motion of a particle in the cylinder is given by the
deformation mapping function.  In the inertial frame $S$, using
Cartesian coordinates, the motion of the particle is given by
Eq.~(\ref{defMapFcn}), where the particle in the reference
configuration at time $t=t_o$ has coordinates $Z^m$. The
coordinates $Z^m$ label the particle in the Lagrangean desription.
Using the transformation to the rotating frame in
Eq.~(\ref{rotatingFrameTransformation}), the motion of the
particle with label $Z^m$ with respect to the rotating $S^\prime$
frame Cartesian coordinates is given by
\begin{equation}\label{DefMapFnPrime}
z^{\prime \, k}(Z^m,t) = A_{kj}(t) \,  z^j(Z^m,t)
\end{equation}

In discussing the transformation to the rotating system of
coordinates, I must distinguish between the velocity of a given
particle in the medium (the Lagrangean picture) and the velocity
field (the Eulerian picture). The Lagrangean velocity ${\rm v}_i
(S;Z^m;t)$ of a given particle (whose coordinates in the reference
configuration are $Z^m$) with respect to the inertial frame $S$ is
defined as the partial time derivative of that particle's
$z^i$-coordinates, when holding $Z^m$ constant
\begin{equation}\label{velocityParticle}
{\rm v}_i (S;Z^m;t) = \frac{\partial z^i(Z^m,t)}{\partial t}
\end{equation}
The Eulerian velocity field, $v_i(z^k,t)$, with respect to the frame $S$ is a
function of coordinates $z^k$ and time $t$ and is related to the
Lagrangean (particle) velocity by
\begin{equation}
{\rm v}_i (S;Z^m;t)  =  {\rm v}_i(S;Z^m(z^k;t),t) = v_i(z^k,t) \label{velocity2}
\end{equation}
where I used Eq.~(\ref{defMapFcnInverse}) to
express the particle coordinate $Z^m$ in terms of its position
$z^k$ at time $t$.

I can describe the same particle's velocity (whose coordinates in
the (inertial frame) reference configuration are $Z^m$) with
respect to the rotating frame of reference $S^\prime$.  The
velocity of this particle with respect to the rotating frame
$S^\prime$ is
\begin{equation}\label{velocityParticlePrime}
{\rm v}^\prime_i (S^\prime;Z^m;t) = \frac{\partial z^\prime_i(Z^m,t)}{\partial
t} = \frac{\partial }{\partial t} \, \left[ \, A_{ik}(t) \,
z_k(Z^m,t) \,  \right] =
\dot{A}_{ik}(t) \, z_k(Z^m,t) +
A_{ik}(t) \, {\rm v}_k (S;Z^m;t)
\end{equation}
where I used Eq.~(\ref{rotatingFrameTransformation}) to express
the particle's coordinates in terms of coordinates in the $S$
frame and the dot on $\dot{A}_{ik}$ indicates differentiation with
respect to time.

The components ${\rm v}_i (S;Z^m;t)$ and ${\rm v}^\prime_i
(S^\prime;Z^m;t)$ represent physically distinct vectors (geometric
objects). Each of these vectors can be expressed on the other
basis.   In particular, according to the standard transformation
rules for vector components, I have
\begin{eqnarray}
{\rm v}^\prime_i (S^\prime;Z^m;t) &  = &  A_{im} \, {\rm v}_m (S^\prime;Z^m;t) \label{velocityA} \\
{\rm v}^\prime_m (S;Z^m;t)  & = &  A_{mi} \, {\rm v}_i(S;Z^m;t) \label{velocityB}
\end{eqnarray}
Equations~(\ref{velocityA}) and (\ref{velocityB}) are the standard
tensor transformation rules for vector components under the coordinate
transformation given in Eq.~(\ref{rotatingFrameTransformation}).
In summary, I must distinguish between four (Cartesian component)
velocities~\cite{SyngeSchild1981}: \newline

\noindent ${\rm v}_i (S;Z^m;t) = \dot{z}_i(Z^m,t) $  = components on $z$-axes
of particle velocity with respect to $S$ \\

\noindent ${\rm v}_i (S^\prime;Z^m;t)$  =
components on $z$-axes of particle velocity with respect to $S^\prime$ \\

\noindent ${\rm v}^\prime_i (S^\prime;Z^m;t) = \dot{z}^\prime_i(Z^m,t)$ = components
on $z^\prime$-axes of particle velocity with respect to $S^\prime$  \\

\noindent ${\rm v}^\prime_m (S;Z^m;t) $  = components on $z^\prime$-axes
of particle velocity with respect to $S$ \newline

These four velocities are related. Multiplying
Eq.~(\ref{velocityParticlePrime}) by $A_{im}(t)$, summing over
index $i$ and using the orthogonal matrix properties
\begin{eqnarray}
A_{in}(t) \, A_{im}(t)  &  = & \delta_{nm} \label{ortho1} \\
A_{ni}(t) \, A_{mi}(t)  &  = & \delta_{nm} \label{ortho2}
\end{eqnarray}
leads to~\cite{SyngeSchild1981}
\begin{equation}\label{VelTransformation1}
{\rm v}_n (S^\prime;Z^m;t) = {\rm v}_n (S;Z^m;t) +
\omega_{nk}(S^\prime,S) \, z_k(Z^m,t)
\end{equation}
where $\omega_{nk}(S^\prime,S)$ is the Cartesian angular
velocity tensor of frame $S^\prime$ with respect to frame $S$:
\begin{equation}\label{angularVel}
\omega_{nk}(S^\prime,S) = A_{in}(t) \, \dot{A}_{ik}(t)
\end{equation}
The tensor $\omega_{nk}(S^\prime,S)$ describes the time-dependent
rotation of frame $S^\prime$ with respect to frame $S$. Similarly,
using Eq.~(\ref{defMapFcn}), substituting the inverse relation
\begin{equation}\label{ParticleCoordTransf}
z_i(Z^m,t) = A_{ni}(t) \, z^\prime_n(Z^m,t)
\end{equation}
carrying out the time differentiation, multiplying by $A_{mi}$,
summing over $i$ and use of the orthogonality relations in
Eq.~(\ref{ortho2}) leads to
\begin{equation}\label{VelTransformation2}
{\rm v}^\prime_n (S;Z^m;t) = {\rm v}^\prime_n (S^\prime;Z^m;t) +
\omega_{kn}^\prime(S^\prime,S) \, z^\prime_k(Z^m,t)
\end{equation}
where the angular velocity tensor components are expressed with
respect to the $S^\prime$ frame Cartesian basis:
\begin{equation}\label{OmegaTransfRule}
\omega^\prime_{nm}(S^\prime,S) \, = \, A_{mi} \, \dot{A}_{ni} \, \,
= A_{ni} \, A_{mk} \,  \omega_{ik}(S^\prime,S)
\end{equation}
Equation (\ref{VelTransformation1}) and (\ref{VelTransformation2})
are the well-known  rules for transformating {\it particle}
velocity to a rotating frame of reference~\cite{SyngeSchild1981}.

Next, I derive the equation that relates the Cartesian components
of the velocity field in $S$, $v_i(z^k,t)$, to the velocity field
in $S^\prime$, $v^\prime_i(z^{\prime \, k},t)$.
Equation (\ref{velocity2}) relates the velocity field in the $S$
frame to the Lagrangean (particle) velocity. Similarly, the
velocity field with respect to the $S^\prime$ frame is given by
\begin{equation}\label{eqVelPrime}
v^\prime_i(z^{\prime \, k},t)= {\rm v}^\prime_i(S^\prime;Z^m;t)
= \frac{\partial \, z^{\prime \, i}(Z^m,t)}{\partial t} =
{\rm v}^\prime_i(S^\prime;Z^m(z^k,t);t) =
{\rm v}^\prime_i(S^\prime;Z^m(A_{nk} \, z^{\prime \, n},t);t)
\end{equation}
where I used the inverse relation $z^k =A_{nk} \, z^{\prime \,
n}$. Using Eq.~(\ref{velocity2}) and (\ref{eqVelPrime}) in the
left and right most terms in Eq.~(\ref{velocityParticlePrime}), I
obtain a relation between the velocity fields in frames $S$ and
$S^\prime$
\begin{equation}\label{VelFieldTransformation}
v^{\prime}_i(z^{\prime \, k},t) = \dot{A}_{ik}(t) \, z_k + A_{ik}(t) \, v_k(z^n,t)
\end{equation}
In Eq.~(\ref{VelFieldTransformation}), the coordinates $z^{\prime
\, k}$ and $z^n$ are related by
Eq.~(\ref{rotatingFrameTransformation}). Multiplying
Eq.~(\ref{VelFieldTransformation}) by $A_{im}$, summing over
index, $i$, using the  inverse transformation in
Eq.~(\ref{ParticleCoordTransf}) and
\begin{equation}\label{OmegaInverseTrans}
\omega_{jk}(S^\prime,S)=A_{mj} \, A_{nk} \, \omega^\prime_{mn}(S^\prime,S)
\end{equation}
leads to
\begin{equation}\label{FinalVelocityTrans}
v_j(z_m,t)=A_{ij}(t) \, v^\prime_i(z^{\prime}_n,t) -A_{kj}(t)
\, \omega^\prime_{kn}(S^\prime,S)\,z^{\prime}_n
\end{equation}
Eq.~(\ref{FinalVelocityTrans}) is the desired  rule for
transformation of the Eulerian velocity field from the inertial
frame $S$ to the rotating frame $S^\prime$. Note that the right
side of Eq.~(\ref{FinalVelocityTrans}) depends only on $S^\prime$
frame coordinates $z^\prime_n$ and the left side depends on $S$
frame coordinates $z_m$.  Furthermore, the velocity field
components on the left side of Eq.~(\ref{FinalVelocityTrans}) are
taken on the (Cartesian) inertial frame $S$ basis vectors, and on
the right side all components are expressed on the (Cartesian)
rotating frame $S^\prime$ basis vectors.

\subsection{Dynamical Equation in the Rotating Frame}

In what follows, I transform the momentum balance
Eq.~(\ref{stressEq}) to the corotating system of coordinates
$S^\prime$.  For simplicity, I do this transformation using
Cartesian coordinates for both the inertial frame $S$ and
corotating frame $S^\prime$.  I use the transformation of the
velocity field given in Eq.~(\ref{FinalVelocityTrans}) to compute
the terms that appear in Eq.~(\ref{stressEq}).    Taking the gradient of
the velocity in Eq.~(\ref{FinalVelocityTrans})
\begin{equation}\label{VelocityGradient}
\frac{\partial v_j(z_m,t)}{\partial \, z_k} =
A_{nk} \, A_{ij} \ \frac{\partial
v^\prime_i(z^\prime_n,t)}{z^\prime_n}-
A_{mk}\, A_{nj} \, \omega^\prime_{nm}(S^\prime,S)
\end{equation}
where I used the chain rule for differentiation
$\frac{\partial}{\partial z_k}= A_{mk} \frac{\partial}{\partial \,
z^\prime_m}$ since $z_k$ and $z^\prime_m$ are related by Eq.
(\ref{DefMapFnPrime}).  Next, I compute the time derivative of the
velocity that occurs in Eq.~(\ref{stressEq}) and  I express the
right side in terms of $S^\prime$ frame components and
coordinates:
\begin{equation}\label{VelocityTimeDerivative}
\frac{\partial v_i(z_m,t)}{\partial \, t} = \dot{A}_{mi} \,
v^\prime_m + A_{li} \, \omega^\prime_{nm} \, z^\prime_m \,
\frac{\partial v^\prime_l}{\partial \, z^\prime_n}
+ A_{mi} \, \frac{\partial v^\prime_m}{\partial \, t} -
\dot{A}_{ri} \, \omega^\prime_{rn} \, z^\prime_n -
A_{mi} \, \dot{\omega}^\prime_{mn} \, z^\prime_n - A_{li} \,
\omega^\prime_{nm} \, \omega^\prime_{ln} \, z^\prime_m
\end{equation}
where I have omitted the frame labels of the angular velocity and
the coordinate arguments in the velocity.  The divergence of the
stress transforms as a vector under the transformation to the
rotating frame
\begin{equation}\label{stressTrans}
\frac{\partial \, \sigma_{ik}}{\partial \, z_k}=
A_{ai} \, \frac{\partial \, \sigma^\prime_{ab}}{\partial \, z^\prime_b}
\end{equation}

Substituting Eqs.~(\ref{FinalVelocityTrans})--(\ref{stressTrans})
into the inertial-frame momentum balance Eq.~(\ref{stressEq}) and
simplifying, leads to the dynamical equation for the (Cartesian)
stress tensor in the rotating frame $S^\prime$
\begin{eqnarray}
\frac{1}{\rho} \frac{\partial \sigma^\prime_{ik}}{\partial z^\prime_k}
= \frac{\partial v^\prime_i}{\partial t} + v^\prime_n \frac{\partial
v^\prime_i}{\partial z^\prime_n}
+ \left(  \omega^\prime_{im} \, \omega^\prime_{mn} -\frac{\partial \,
{\omega}^\prime_{in}}{\partial \, t}
\right) \, z^\prime_n
-2 \, \omega^\prime_{im} \, v^\prime_m
\label{corotatingStresEq}
\end{eqnarray}
where $\sigma^{\prime}_{ik}$ are the stress components in
Cartesian coordinates $z^\prime_k$, $v^\prime_n$ is the Eulerian
velocity field that depends on $z^\prime_k$ and $t$, and the
(Cartesian) components of the angular velocity tensor  in
$S^\prime$ are given by Eq.~(\ref{OmegaTransfRule}). In Eq.\
(\ref{corotatingStresEq}), all repeated subscripts are summed.
Note that all velocities that appear in Eq.\
(\ref{corotatingStresEq}) refer to the corotating frame $S^\prime$
and that all tensor components are taken on the $S^\prime$ frame
Cartesian basis vectors. The first two terms in Eq.\
(\ref{corotatingStresEq}) are the acceleration (including the
convective term) as seen in the corotating system of coordinates.
The third term $\omega^\prime_{im}\, \omega^\prime_{mn} \,
z^\prime_n$  is the centrifugal acceleration.  The fourth term,
$\dot{\omega}^\prime_{in} \, z^\prime_n$ is the angular
acceleration.  The last term, $-2 \, \omega^\prime_{in} \,
v^\prime_n$ is the Coriolis acceleration.

The dynamical Eq.~(\ref{corotatingStresEq}) is a tensor equation;
the quantities $\sigma^\prime_{kn}$, $\omega^\prime_{mn}$, and
$v^\prime_n$, are Cartesian tensors.  Under orthogonal
transformations from one Cartesian system to another,
Eq.~(\ref{corotatingStresEq}) is covariant: it has the same form.
The group of symmetry operations may be extended to
transformations between curvilinear coordinates by writing Eq.\
(\ref{corotatingStresEq}) in a manifestly covariant form as:
\begin{equation} \label{CylindricalCorotatingStresEq}
\frac{1}{\rho} \, \, \tilde{\sigma}^{ik}_{~~ \, ; k} =
\frac{\partial \tilde{v}^i}{\partial t}
+\tilde{v}^{n} \, \tilde{v}^{i}_{~ ; n}
+ 2 \, \tilde{\omega}_n^{\,~ \,i} \, \tilde{v}^n
+ \left(
\tilde{\omega}_{m}^{~~ \, i} \, \tilde{\omega}_{n}^{~ \, m}
  + \frac{\partial \tilde{\omega}_{n}^{~~ i}}{\partial \, t}
 \right) \, \tilde{\zeta}^n
\end{equation}
where the tilde over each tensor indicates that the components are
taken on generalized curvilinear coordinate (such as cylindrical)
basis vectors in the rotating frame of reference $S^\prime$.
Equation~(\ref{CylindricalCorotatingStresEq}) is general; it is
valid for all motions and all materials.
Equation~(\ref{CylindricalCorotatingStresEq}) expresses Newton's
law for a continuous medium in an arbitrary rotating frame of
reference that is defined by a general (Cartesian) angular
velocity tensor $\omega_{ik}(S^\prime,S)=A_{li}(t) \,
\dot{A}_{lk}(t)$, which relates the inertial frame $S$ and
rotating frame $S^\prime$, with coordinates related by
Eq.~(\ref{rotatingFrameTransformation}). The quantities
$\tilde{\zeta}^n$ are the contravariant components of the position
vector in curvilinear coordinates $x^{\prime \, m}$, which are
related to the Cartesian position vector components $z^{\prime \,
k}$ by
\begin{equation}\label{positionVectorComponents}
\tilde{\zeta}^m =
\frac{\partial x^{\prime m}}{ \partial z^{\prime k}} \, \, z^{\prime k}
\end{equation}
The rotating frame curvilinear components of stress, velocity and
angular velocity, $\tilde{\sigma}^{a b}$, $\tilde{v}^a $, and
$\tilde{\omega}_a^{~ \, b} $, are related to their rotating frame
Cartesian components, $\sigma^\prime_{ij}$, $ v^{\prime k}$, and
$\omega^\prime_{ij}$, by:
\begin{eqnarray}
\tilde{\sigma}^{a b} & = &  \frac{\partial x^{\prime a}}{ \partial z^{\prime i}}
\frac{\partial x^{\prime b}}{ \partial z^{\prime j}}
\, \, \sigma^\prime_{ij}   \label{tensorDefs1}  \\
\tilde{v}^a  &  =  & \frac{\partial x^{\prime a}}{ \partial z^{\prime k}}
\, \, v^{\prime k}          \label{tensorDefs2} \\
\tilde{\omega}_a^{~ \, b}  & = & \frac{\partial z^{\prime i}}{ \partial x^{\prime a}}
\frac{\partial x^{\prime b}}{\partial z^{\prime j}}
 \, \, \omega^\prime_{ij}   \label{tensorDefs3}
\end{eqnarray}
where $z^{\prime k}$ and $x^{\prime a}$ are the Cartesian and
curvilinear coordinates in the rotating frame $S^\prime$. Note
that the partial derivative with respect to coordinates $z^{\prime
k}$ in Eq.~(\ref{corotatingStresEq})  has been replaced by a
covariant derivative with respect to the curvilinear coordinates
$x^{\prime k}$ in Eq.~(\ref{CylindricalCorotatingStresEq}).  See
Tables I and II for summary of the notation. Identification of the
meaning of the various terms in
Eq.~(\ref{CylindricalCorotatingStresEq}), such as the Coriolis
acceleration and centrifugal accelleration is clear from
comparison with Eq.~(\ref{corotatingStresEq}).

\section{Rotating Cylinder Equations in Corotating Coordinates}

In order to obtain the explicit equations for a rotating cylinder
from Eq.~(\ref{CylindricalCorotatingStresEq}), I need to compute
$\tilde{v}^n$, $\tilde{\omega}_m^{~~n}$, and $\zeta^n$.  As
described previously,  I take the reference configuration of the
cylinder to be the stationary cylinder at $t=-\infty$ in the
inertial frame $S$. I assume that the cylinder experiences a slow
angular acceleration (such as given by
Eq.~(\ref{angularAcceleration})) that lasts approximately time $
\Delta t=- 1 / \epsilon $, and has peak magnitude at $t=-T$, with
$1 / \epsilon << T$. Equation~(\ref{CylindricalCorotatingStresEq})
is general; it is valid for all motions and all materials. In what
follows, I restrict my remarks to a perfectly elastic cylinder.

In the inertial frame  $S$, as the cylinder increases its angular
velocity, the material particles of the cylinder move outward in a
spiral path, with some motion in the z-direction. In the inertial
frame $S$, the contravariant components of the velocity field in
cylindrical coordinates are
\begin{equation}\label{velocityFieldInertialCylindrical}
\bar{v}^i = (\bar{v}^1, \omega(t), \bar{v}^3)
\end{equation}
where $\bar{v}^1$ and $\bar{v}^3$ are the radial and
$z$-components of velocity and where the azimuthal component
$\omega(t)$ is given by Eq.~(\ref{angularAcceleration}). I obtain
the velocity components in cylindrical coordinates in the
corotating frame $S^\prime$ as follows:  first, transform
$\bar{v}^i$ to inertial frame Cartesian components $v^i$ using the
standard vector transformation rule between Cartesian and
cylindrical coordinates.  Next, use Eq.~(\ref{FinalVelocityTrans})
to transform the inertial frame Cartesian componets $v^i$ to the
rotating $S^\prime$ frame Cartesian velocity field $v^{\prime \,
i}$.  Finally, use the standard tensor transformation rules,
between Cartesian and cylindrical coordinates (both in the
rotating frame), to transform the velocity field from Cartesian
components  $v^{\prime \, i}$ to cylindrical
(rotating frame) components $\tilde{v}^{\prime \, i}$, where I
made use of the angular velocity components
\begin{equation}\label{angularVelocityTensor}
 \omega_{ik}(S^\prime,S)=A_{li} \, \dot{A}_{lk}= \omega(t)
 \left(
\begin{array}{ccc}
  0 & -1 & 0 \\
  +1 & 0 & 0 \\
  0 & 0 & 0
\end{array} \right)
=\omega^{\prime}_{i k}(S^\prime,S) = A_{im} \, A_{kn} \, \omega_{mn}(S^\prime,S)
\end{equation}
Following this procedure, I obtain the velocity components in
cylindrical coordinates in the corotating frame $S^\prime$
\begin{equation}\label{velocityFieldCorotatingCartesian}
\tilde{v}^{i}  = \left( \bar{v}^1, \, 0, \, \bar{v}^3 \,
\right)
\end{equation}
where the azimuthal component of velocity is zero for all time, by
construction of the corotating frame $S^\prime$, as expected.

In the corotating frame $S^\prime$, there is particle motion
around the time $t\approx -T$.  However, at $t=0$ the particles
have reached their new (deformed) steady-state positions and
motion has ceased; the velocity field is given by
\begin{equation}\label{ZeroVelocityField}
\tilde{v}^{k}=(0,0,0)
\end{equation}
which is the velocity field of a rigid body.  Since the velocity
field is zero at $t=0$, the Coriolis acceleration term in
Eq.~(\ref{CylindricalCorotatingStresEq}) does not contribute in
steady-state rotation.

Using Eq.(\ref{tensorDefs3}), the angular velocity tensor in
cylindrical coordinates in the corotating frame $S^\prime$ is
\begin{equation}\label{angularVelocityTensorCorotFrame}
\tilde{\omega}_a^{~ \, b} = \left( \begin{array}{ccc}
  0 & \omega(t)/r^\prime & 0   \\
  -r^\prime \omega(t) & 0 & 0 \\
   0 & 0 & 0
\end{array}  \right)
\end{equation}
At $t=0$, the time derivative of the angular velocity tensor is
zero. Using this fact, and Eq.~(\ref{ZeroVelocityField}) and
(\ref{angularVelocityTensorCorotFrame}), the stress
Eqs.~(\ref{CylindricalCorotatingStresEq}) in the corotating frame
are given by
\begin{eqnarray}
\tilde{\sigma}^{11}_{~~ ,1} + \tilde{\sigma}^{13}_{~~ ,3} +
\frac{\tilde{\sigma}^{11}}{r^\prime}- r^\prime \tilde{\sigma}^{22} & = &  -
\rho r^\prime \omega_o^2   \label{stress7} \\
\tilde{\sigma}^{12}_{~~ ,1}  + \tilde{\sigma}^{23}_{~~ ,3} +
\frac{3}{r^\prime} \tilde{\sigma}^{12}  & = &   0   \label{stress8} \\
\tilde{\sigma}^{13}_{~~ ,1} + \tilde{\sigma}^{33}_{~~ ,3} +
\frac{\tilde{\sigma}^{13}}{r^\prime}   & = &  0  \label{stress9}
\end{eqnarray}
Equations~(\ref{stress7})--(\ref{stress9}) are the equations
satisfied by the stress tensor in cylindrical components at $t=0$,
in steady-state rotation in the corotating frame $S^\prime$. In
terms of physical components (see Eq.~(\ref{pc1})--(\ref{pc6})),
Eq.~(\ref{stress7})--(\ref{stress9}) become:
\begin{eqnarray}
\frac{\partial \tilde{\sigma}^{r r}}{\partial r^\prime} + \frac{\partial \tilde{\sigma}^{r z}}{\partial z^\prime} +
\frac{\tilde{\sigma}^{rr} - \tilde{\sigma}^{\phi \phi}}{r^\prime} & = &  -
\rho \, r^\prime \omega_o^2   \label{stP4} \\
\frac{\partial}{\partial r^\prime} \left( \frac{1}{r^\prime} \tilde{\sigma}^{r \phi} \right)
+ \frac{1}{r^\prime} \frac{\partial \tilde{\sigma}^{\phi z}}{\partial z^\prime}  +
\frac{3}{r^{\prime \, 2}} \, \, \tilde{\sigma}^{r \phi}  & = &   0   \label{stP5} \\
\frac{\partial \tilde{\sigma}^{r z}}{\partial r^\prime}  +
\frac{\partial \tilde{\sigma}^{z z}}{\partial z^\prime}  +
\frac{\tilde{\sigma}^{r z}}{r^\prime}   & = &  0  \label{stP6}
\end{eqnarray}

Note that Eq.~(\ref{stP4})--(\ref{stP6}) in the corotating frame
$S^\prime$ have the same form as Eq.~(\ref{st4})--(\ref{st6}) in
the inertial frame $S$. However, the key point is that the
corotating frame Eq.~(\ref{stP4})--(\ref{stP6}) have a distinct
advantage: the quadratic strain gradient terms in the definition
of the strain in Eq.~(\ref{strainDisplacement}) can be dropped
because they are small in the corotating frame $S^\prime$.  The
same is not true in the inertial frame $S$.

To proceed with the solution in the corotating frame, the
constitutive Eq.~(\ref{constitutiveEq}) (in inertial frame $S$)
must be transformed to the corotating frame $S^\prime$ by taking
cylindrical components in the $S$ frame and using
Eq.~(\ref{SandSprime}) to obtain an expression of the same form as
in  Eq.~(\ref{constitutiveEq}) but in the corotating frame
$S^\prime$. In this transformation, the Lam\'{e} constants are
treated as invariants.  So the constitutive relations in the
corotating frame $S^\prime$ are the same as in the inertial frame
$S$.

In the corotating frame $S^\prime$,  dropping the quadratic terms
in displacement gradients, which relate the displacement field to
strain, is justified since in the rotating frame $S^\prime$ these
terms can be considered small for moderate angular velocity.
Therefore, the solution in the corotating frame $S^\prime$ can
proceed in an analogous way to that of the `standard method', but
I have not made the (incorrect) approximation of dropping
quadratic displacement gradient terms in the inertial frame.

The stress tensor is objective, so the stress in the rotating
frame has the same meaning as in the inertial frame, see
Eq.~(\ref{stressTransformation}). The boundary conditions on the
stress tensor components in the corotating frame are the same as
in the inertial frame, due to the objectivity of stress tensor.
Alternatively, one can verify that the boundary conditions on the
stress in the rotating frame are the same as in the inertial
frame~\cite{bahdersBoundaryConditionsNotes}.

\section{Plane Stress Solution}

The solution of the problem of stress in a rotating cylinder in
the corotating frame of reference now follows.   The solution in
the corotating frame parallels the solution in the `standard
method'~\cite{Love1944,LLelasticity1970,Nadai1950,Sechler1952,Timoshenko1970,VolterraGaines1971},
except that the incorrect approximation of dropping the quadratic
strain gradient terms in the inertial frame is avoided.

I assume that in the inertial frame $S$, the cylinder is rotating
at angular velocity $\omega_o$ and has radius $b$. Under the
assumption of plane stress~\cite{Timoshenko1970}, where
\begin{equation} \label{PlaneStress}
\tilde{\sigma}^{z z} = \tilde{\sigma}^{\phi z} =
\tilde{\sigma}^{r z} =0
\end{equation}
with boundary condition of zero stress on the long peripheral
surface:
\begin{equation}\label{bc}
\tilde{\sigma}^{r r}  \vert_{r^\prime = b} =0
\end{equation}
As mentioned above, stress is an objective
tensor~\cite{Eringen1962,Narasimhan1992}, so the physical meaning
of the boundary conditions in Eq.~(\ref{PlaneStress}) and
(\ref{bc}) in the corotating frame $S^\prime$ are the same as the
physical meaning of the analogous conditions in the inertial frame
(as used, for example by Timoshenko~\cite{Timoshenko1970}). Also,
as shown above,  I may take the (transformed) linear elastic
relations in the corotating frame to be of the same form as in the
inertial frame:
\begin{eqnarray}
\tilde{e}^{rr} & = & \frac{1}{E} \, \left( \tilde{\sigma}^{rr} - \nu \, \tilde{\sigma}^{\phi\phi}
\right) \label{elastic1} \\
\tilde{e}^{\phi \phi} & = & \frac{1}{E} \, \left( \tilde{\sigma}^{\phi\phi} - \nu \, \tilde{\sigma}^{rr}
\right) \label{elastic2}
\end{eqnarray}
where Young's modulus $E$ and Poisson's ratio $\nu$ are related to
the Lam\`{e} constants by
\begin{eqnarray}
E & = & \frac{\mu \left( 3 \lambda + 2 \mu\right)}{\lambda+\mu} \label{EDef} \\
\nu & = & \frac{ \lambda} {2\left( \lambda+\mu \right)}
\label{nuDef}
\end{eqnarray}
where the Lam\`{e} constants are treated as invariant scalars in
the transformation. The elastic relations in Eq.~(\ref{elastic1})
and (\ref{elastic2}) are in the corotating frame $S^\prime$.  They
can be obtained from the linear elastic relations in the inertial
frame $S$, Eq.~(\ref{constitutiveEq}), by using the
transformations to the corotating frame in
Eq.~(\ref{stressTransformation})  and
(\ref{strainTransformationGeneral}), and the coordinate
transformation in Eq.~(\ref{CylTrans}). In doing the
transformation of the elastic relations to the corotating frame, I
am assuming that the tensor that enters in the inertial frame
elastic relations in Eq.~(\ref{constitutiveEq}) is the Eulerian
strain tensor given in Eq.(\ref{strainDisplacement}) and the
quadratic terms have not been dropped. (As discussed earlier,
dropping the quadratic terms in the displacement gradients is done
in the corotating frame.)

For small displacements in corotating frame $S^\prime$, the
gradients of the displacement vector in frame $S^\prime$ can be
assumed small---for moderate angular velocity $\omega_o$, so the
quadratic terms in the gradient of the displacement can be
neglected.  Therefore, in the corotating frame $S^\prime$, I take
the relation between the radial component of the displacement
vector, $\tilde{u}$, and the physical components of strain,
$\tilde{e}_{rr}$ and $\tilde{e}_{\phi \phi}$, to be  (compare with
Eq.~(\ref{strainU1}) and (\ref{strainU2}))
\begin{eqnarray}\label{StressStrainCorotating}
\tilde{e}_{rr} & = &  \frac{\partial \tilde{u}}{\partial r^\prime} \label{strain1Rot} \\
\tilde{e}_{\phi\phi} & = &  \frac{\tilde{u}}{ r^\prime} \label{strain2Rot}
\end{eqnarray}
where the tilde on $\tilde{u}$ indicates that the radial component of
displacement field is taken in the corotating frame $S^\prime$ and
the prime on $r^\prime$ indicates that the cylindrical radial coordinate is in
the corotating frame $S^\prime$, see the transformation in Eq.~(\ref{CylTrans}).

Substituting Eq.~(\ref{strain1Rot}) and (\ref{strain2Rot}) into
Eq.~(\ref{elastic1}) and (\ref{elastic2}), leads to relations
between the physical components of stress and the radial
displacement field
\begin{eqnarray}
\tilde{\sigma}^{rr} & = &  \frac{E}{1-\nu^2} \,
\left( \frac{\partial \tilde{u}}{\partial \, r^\prime}
+ \nu \, \frac{\tilde{u}}{r^\prime}\right)
\label{StressDisplacementRR} \\
\tilde{\sigma}^{\phi\phi} & = &  \frac{E}{1-\nu^2} \,
\left( \frac{\tilde{u}}{r^\prime}
+ \nu \,  \frac{\partial \tilde{u}}{\partial \, r^\prime} \right)
\label{StressDisplacementPHIPHI}
\end{eqnarray}
Substituting these relations  into Eq.~(\ref{stP4}) leads to a
differential equation for the displacement in the rotating frame
$S^\prime$
\begin{equation}\label{uEqn}
r^{\prime \, 2} \frac{\partial ^2 \tilde{u}}{\partial r^{\prime \, 2}}
+ r^{\prime} \frac{\partial \tilde{u}}{\partial r^{\prime }} -
\tilde{u} = -  \frac{1-\nu^2}{E} \, \rho \, \omega_o^2 \,
r^{\prime \, 3}
\end{equation}
The general solution is~\cite{Timoshenko1970}
\begin{equation}\label{uSolution}
\tilde{u} = \frac{1}{E} \, \left[ (1-\nu) \, C_1 \, r^\prime -
(1+\nu) \, C_2 \, \frac{1}{r^\prime} - \frac{1-\nu^2}{8}\, \rho \,
\omega^2_o \, r^{\prime \, 3} \right]
\end{equation}
Substitution of this solution into
Eq.~(\ref{StressDisplacementRR})and
(\ref{StressDisplacementPHIPHI}) I obtain
\begin{eqnarray}
\tilde{\sigma}^{rr} & = &  C_1 + C_2 \, \frac{1}{r^{\prime \, 2}} - \frac{3+\nu}{8}
\, \rho \, \omega^2_o \, r^{\prime \, 2}
\label{StressDispRR} \\
\tilde{\sigma}^{\phi\phi} & = &   C_1 - C_2 \, \frac{1}{r^{\prime \,
2}} - \frac{1+3\nu}{8} \,\rho \, \omega^2_o \, r^{\prime \, 2}  \label{StressDispPHIPHI}
\end{eqnarray}
The stresses at $r^\prime=0$ must remain finite, so I take
$C_2=0$.  Applying the boundary condition on the long peripheral
surface, Eq.~(\ref{bc}) leads to $C_1=(3+\nu) \, \rho \,\omega^2_o
\, b^2  / 8$ and the stresses
\begin{eqnarray}
\tilde{\sigma}^{rr} & = &  \frac{3+\nu}{8}\rho
\, \omega_o^2 \, \, (b^2 - r^{' \, 2}  )  \label{stressRR} \\
\tilde{\sigma}^{\phi \phi}  & = & \frac{1}{8} \, \rho \, \omega_o^2 \left[
(3+\nu)
b^2 - (1+ 3\nu) r^{\prime \, 2} \right]   \label{stressFiFi}
\end{eqnarray}
The physical stress components in Eq.~(\ref{stressRR}) and
(\ref{stressFiFi}) are in the corotating frame $S^\prime$.
However, due to the transformation between the corotating frame
and the inertial frame in Eq.~(\ref{stressTransformation}), and
the coordinate transformation in Eq.~(\ref{CylTrans}), the
corotating frame components in Eq.~(\ref{stressRR}) and
(\ref{stressFiFi}) are equal to the inertial frame components of
stress.   Using the expressions in the rotating frame, such as
Eq.~(\ref{stP4})---(\ref{stP6}), expressions for plane strain and
other boundary conditions can be derived for rotating cylinders,
disks and annular rings, see
Ref.~\cite{Love1944,LLelasticity1970,Nadai1950,Sechler1952,Timoshenko1970,VolterraGaines1971}.

\section{Summary}

The classic problem of stress in rotating disks or cylinders is
important in applications to turbines, generators, and whenever
large rotational speeds exist.  The textbook problem of stress in
perfectly elastic disks or cylinders is solved in standard
texts~\cite{Love1944,LLelasticity1970,Nadai1950,Sechler1952,Timoshenko1970,VolterraGaines1971}.
The `standard method' of solution begins with Eq.~(\ref{stressEq})
and drops terms that are quadratic in strain gradient in the
definition of the strain, see Eq.~(\ref{strainDisplacement}).
Equation~(\ref{stressEq}) is valid only in an inertial frame of
reference, since it is derived from Newton's second law of motion,
which itself is only valid in an inertial reference frame.

In this work, I have shown that dropping the terms quadratic in
the displacement gradient (in Eq.~(\ref{strainDisplacement})) is
incorrect in the inertial frame in which Eq.~(\ref{stressEq}) is
applied in the `standard method' of
solution~\cite{Love1944,LLelasticity1970,Nadai1950,Sechler1952,Timoshenko1970,VolterraGaines1971}.
I provide an alternative formulation of the rotating elastic
cylinder problem in a frame of reference that is corotating with
the cylinder.  In this corotating frame, I derive the dynamical
equation for the stress (see Eq.~(\ref{corotatingStresEq}) or
(\ref{CylindricalCorotatingStresEq})) and I show that terms
quadratic in the displacement gradient can be dropped because they
are small (for moderate angular speed of rotation). This analysis
in the corotating frame shows that the `standard method' of
solution~\cite{Love1944,LLelasticity1970,Nadai1950,Sechler1952,Timoshenko1970,VolterraGaines1971}
should be interpreted as being carried out in the corotating frame
of reference of the cylinder.

Furthermore, when stresses are computed in rotating disks or
cylinders composed of materials that have more complex
constitutive equations, such as elastic-plastic or viscoelastic
behavior, one must carefully justify dropping the quadratic terms
in displacement gradients. If dropping these terms cannot be
justified, then the problem can be analyzed in a rotating frame,
using the derived Eq.~(\ref{corotatingStresEq}) or
(\ref{CylindricalCorotatingStresEq}).   Another practical
application of the stress Eq.~(\ref{CylindricalCorotatingStresEq})
in the rotating frame is to study elastic waves in bodies during
rotation, where coriolis effects may play a role.

\begin{acknowledgements}
The author thanks Dr. W. C. McCorkle, U.\ S.\ Army Aviation and
Missile Command, for suggesting this problem and providing numerous
discussions. The author thanks Howard Brandt for discussions and
pointing out Ref.~\cite{LLelasticity1970}.
\end{acknowledgements}

\appendix
\section{Conventions}
I specify tensor components on coordinate (non-holonomic) basis
vectors using numerical indices, 1,2,3, such as
$\bar{\sigma}^{12}$. For physical components, which have the
dimensions associated with that quantity, I use lettered indices,
such as $\bar{\sigma}^{r \phi}$. In addition, I must distinguish
between four coordinate systems: Cartesian and cylindrical
coordinates in the inertial frame $S$ and Cartesian and
cylindrical in the corotating frame $S^\prime$.  I use
$z^k=(x,y,z)$ and $x^k=(r,\phi,z)$ for Cartesian and cylindrical
coordinates in inertial frame $S$, respectively.  In corotating
frame $S^\prime$, I use $z^{\prime \,
k}=(x^\prime,y^\prime,z^\prime)$ and $x^{\prime \,
k}=(r^\prime,\phi^\prime,z^\prime)$ for Cartesian and cylindrical
coordinates, respectively. For distinguishing components in these
four coordinate systems, I use an additional mark as follows:
absence of mark and a bar, for Cartesian and cylindrical
components in inertial frame $S$, respectively.  For components in
the corotating frame $S^\prime$, I use a prime and a tilde, for
Cartesian and cylindrical components, respectively. See Table I
and II.


\begin{table}
\caption{Coordinates \label{coordinates}}
\begin{tabular}{cccc}
              & Inertial Frame $S$  &  Rotating Frame $S^\prime$   \\ \tableline
Cartesian     &     $z^k=z_k$        &   $z^{\prime k}=z^\prime_k$
\\ Cylindrical   &     $x^k$        &   $x^{\prime k}  $    \\
\end{tabular}
\end{table}

\begin{table}
\caption{Tensor Components \label{TensorComponents}}
\begin{tabular}{cccc}
              & Inertial Frame  $S$  &  Rotating Frame $S^\prime$ \\ \tableline
Cartesian     &     $\sigma^{ab}(z^k)=\sigma_{ab}$,  $e_{ik}$, $\omega_a^{~ b}$ &
$\sigma^{\prime ab}(z^{\prime k})=\sigma^\prime_{ab}$,  $e^\prime_{ik}$, $\omega^{\prime ~b}_{a}$      \\
Cylindrical   &  $\bar{\sigma}^{ab}(x^k)$, $\bar{e}_{ik}$, $\bar{\omega}^{~ b}_{a}$  &
$\tilde{\sigma}^{ab}(x^{\prime k})$, $\tilde{e}_{ik}$,  $\tilde {\omega}^{~b}_{a}$    \\
\end{tabular}
\end{table}


\begin{thebibliography}{}
\bibitem{FOOTNOTE-1}For the limiting case of a rotating
rigid body, the strain $e_{ik}=0$ because the linear displacement
gradient terms $u_{i;k} + u_{k;i}$  cancel the quadratic terms
$u_{m;i} u^m_{~~ ; k}$, so that the full strain tensor $e_{ik}=0$.
\end{thebibliography}
\end{document}